\documentclass{osa-article}
\usepackage{epstopdf}
\usepackage{placeins}
\let\vec=\mathbf

\usepackage[labelformat=empty]{caption}

\newcommand{\blue}[1]{{\color{black}#1}}

\journal{josab}

\articletype{Research Article}

\begin{document}

\title{Second-harmonic generation in dielectric nanoparticles with different symmetries}

\author{Kristina Frizyuk,\authormark{1}}

\address{\authormark{1}ITMO University, 197101 Saint Petersburg, Russia }

\email{\authormark{*}k.frizyuk@metalab.ifmo.ru} 



\begin{abstract}
In this work we study second-harmonic generation a monocrystalline nanoparticle with a non-centrosymmetric crystalline lattice.
    It was shown that breaking the symmetry of the nanoparticle's shape can significantly affect the second harmonic radiation pattern. We propose a method for explaining and predicting the generated field for arbitrary nanoparticles and provide selection rules for nanoparticles with several different symmetries. \end{abstract}

\section{Introduction}

        Second-harmonic generation (SHG) by resonant nanoparticles has recently been actively studied both theoretically \cite{bib:dean, bib:kivsh0, Wunderlich:13, Bachelier:08,  Bernasconi:1616} and experimentally \cite{Gili:16, bib:gvg4, bib:gvg21, bib:gvg_exp3, bib:gvg1, Baumner:10} in order to develop nanosized light sources.
 {The} absence of phase matching conditions at the subwavelength scales results in significant drop of the generation efficiency, {making the exploitation of resonances in such nanoscale structures the only way to enhance SHG}. {During the last few decades, metallic nanostructures supporting localized surface plasmon esonances have been actively studied~--- while the lattice of typical plasmonic
    materials has a center of inversion, second harmonic (SH) generation is still possible by surface and nonlocal volume effects\cite{Mochan, bib:gvg2, bib:gvg6, bib:gvg_cores, bib:gvg_mult}.}
  {In more recent developments, nanoparticles from dielectrics and semiconductors with a non-centrosymmetric crystalline lattice and bulk second order non-linearity \cite{bib:gvg_exp, bib:gvg_mult2, bib:gvg_bic, doi:10.1021/acs.nanolett.8b00830} have been extensively studied.} Mie-resonances {supported by these dielectric nanoparticles}\cite{bib:Boren} {can} lead to several orders of magnitude increase {in} SHG efficiency compared to metal structures\cite{bib:kivsh}.
{Despite the large number of studies in this area, the full physical picture explaining SHG enhancement in all-dielectric nanostructures has yet to be identified.}
{S}election rules, i.e. rules that determine the correspondence between the modes excited at the fundamental frequency and the modes at the second harmonic, play one of the most important roles. Selection rules for {SHG} in nanoparticles from materials that do not have bulk
    nonlinearity, such as gold, silicon, etc., were studied in detail in~\cite{bib:Dadap} for spherical particles, and in ~\cite{bib:Finazzi} for particles of other shapes.
{D}espite the many studies devoted to the investigation of SHG from nanoparticles with bulk nonlinearity, the nature
    of the multipole composition of the generated fields, as well as the effect of nanoparticle symmetry on it, has not been studied in detail.
{In most of the works, the results of the SH fields' multipole expansion numerical calculations are presented without an explanation of the physical nature
    of the appearance of certain modes in the spectrum.}
{The goal of this} work is {to determine} the selection rules for SHG by dielectric nanoparticle{s}
    with arbitrary {symmetries} and {with a} nonlinear susceptibility tensor $\hat\chi^{(2)}$.
    In particular, we {will} show that a violation of the symmetry can strongly affect the mode composition and, as a result, the radiation pattern of the SH.
        \begin{table}[h]
    \begin{center}
        \includegraphics[width=0.99\linewidth]{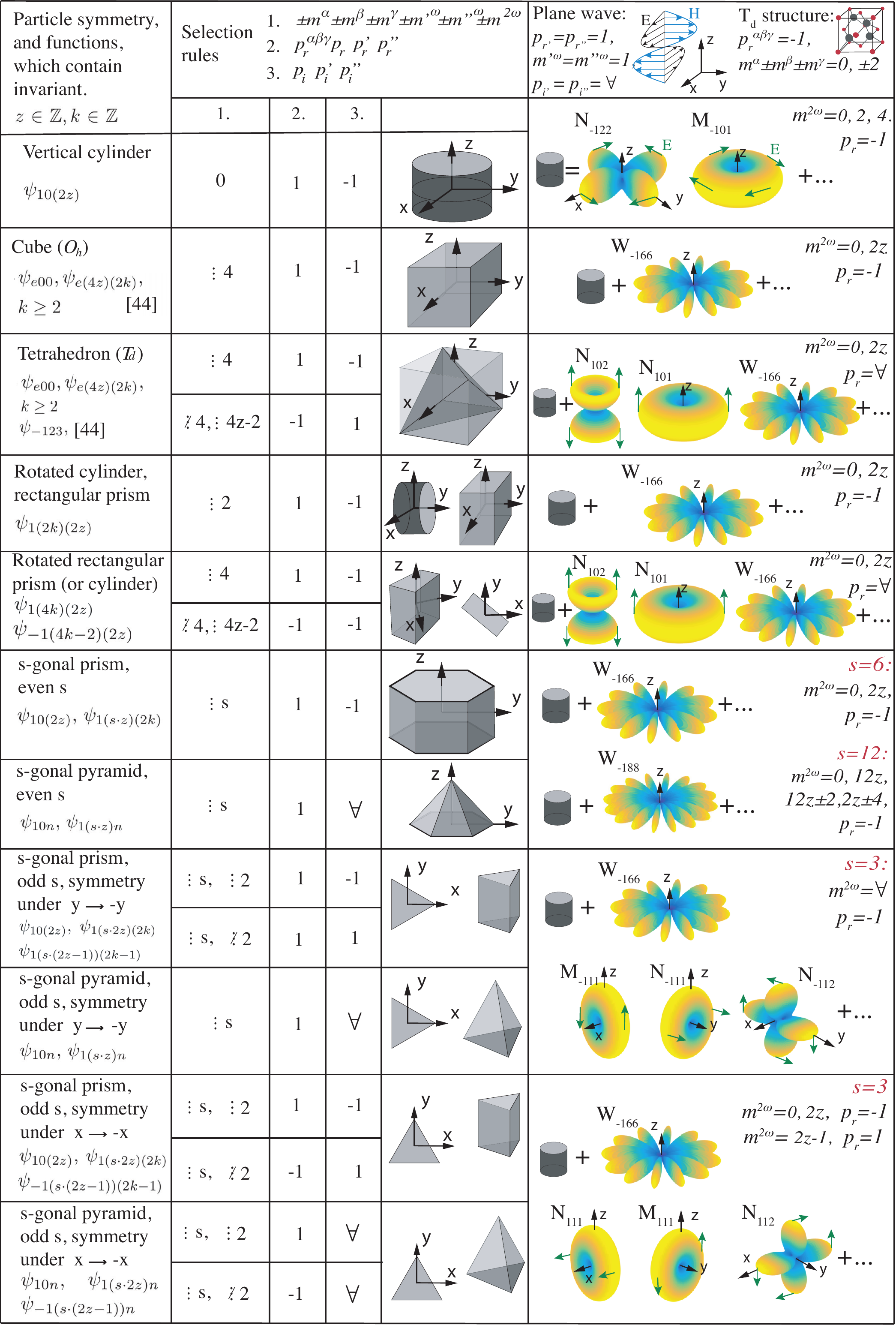}  \caption{Table 1. Selection rules for second harmonic generation in nanoparticles with different shapes \blue{ (left columns), and examples of allowed multipoles on $2\omega$ in GaAs nanostructures under the $x$-polarized plane-wave excitation  (right column).}}
        \label{table1}
        \end{center}
    \end{table} 
 \section{\blue{Derivation of selection rules}}

        According to~\cite{bib:Boren, bib:Stratton, PhysRevB.99.045406, bib:tmatrix, bib:mieharm_fg_main}, {the field} $\vec E$, which {satisfies} the Helmholtz equation
    $\nabla^2 \vec E+ \varepsilon \left(\frac{\omega}{c}\right)^2 \vec E=0$ with frequency $\omega$ in {a} medium with dielectric susceptibility $\varepsilon$,
    can be expanded {into} a series of vector spherical harmonics $\vec N_{p_rmn}(\sqrt{\varepsilon} \frac{\omega}{c},\vec r)$ (electric) and
    $\vec M_{p_rmn}(\sqrt{\varepsilon} \frac{\omega}{c},\vec r)$ (magnetic), introduced in~\cite{bib:arxiv, bib:Boren}. They correspond to the electric field of electric and magnetic multipoles of {the} $n$th order.

{To make further derivations more compact}, the vector spherical harmonics $ \vec M $ and $ \vec N $ will be denoted by the same letter $\vec W_{p_ip_rmn}(\omega)$, {with
    magnetic harmonics \blue{$ \vec M_{p_rmn}$} denoted by index $p_i=(-1)^{n+1}$, and electric harmonics \blue{$ \vec N_{p_rmn}$} by $p_i=(-1)^n$.} Index $n$ corresponds to the order of the multipole, $m$ takes values from
    $0$ to $n$, and $p_r=1$ if the function $\vec W_{p_ip_rmn}(\omega)$ is even with respect t{o r}eflection in the $y=0$-plane ({$\varphi \rightarrow -\varphi$ in spherical coordinates}),
    and $p_r=-1$ if it is odd. The decomposition of the SH field has the form:

    \begin{equation}
        \vec E^{2\omega}(\vec r) =
        \sum_{n=1}^\infty \sum_{m=0}^n\sum_{p_i,p_r} E_0[D_{p_ip_rmn} \vec W^{(3)}_{p_ip_rmn}(2\omega)]
    \end{equation}

    The superscript ${(3)}$ means that the harmonic correspond{s} to {a} diverging wave \cite{bib:Boren}.
    In~\cite{bib:arxiv}, using the multipole expansion of the Green's function \cite{bib:mieharm_fg_main, Yee1994, Tai1972, bib:fg, bib:mieharm_fg} it was shown that the
    coefficients $D_{p_ip_rmn}$ are proportional to the integral of the scalar product of the induced SH polarization
    $\vec P^{2\omega}(\vec r)=\hat\chi^{(2)} \vec E^{\omega}(\vec r) \vec E^{\omega}( \vec r)$ and the vector harmonic with indices ${p_i, p_r, m, n}$
    over the nanoparticle volume:

    \begin{equation}
        D_{p_ip_rmn}\propto \int\limits_V [{\vec{W}}_{p_ip_rmn}(2\omega)\cdot \hat\chi^{(2)} \vec E^{\omega}(\vec r) \vec E^{\omega}( \vec r)]\ dV\:.\label{eq:Dcoeff}
    \end{equation}

    The incident field inside the particle can also be decomposed into vector spherical harmonics \cite{bib:Boren, bib:arxiv},  \blue{where multipolar content depends on the illumination conditions and the nanostructure's symmetry (see, for example, \cite{linearcond})}. The contribution of
    certain multipoles {is usually dominant}, and due to the possibility of selective excitation, we will separately consider the "overlap integrals" of
    three multipoles, two of which relate to the incident field. {Rewriting} the integrand \eqref{eq:Dcoeff} by components and express{ing} the scalar projections of vector spherical harmonics as scalar
    products with cartesian basis vectors we show that the generation of the multipole $\vec W_{p_ip_rm n}$ in the SH from the multipoles
    $\vec W_{p_i'p_r'm'n'}$ and $\vec W_{p_i''p_r''m'' n''}$ in the fundamental {mode} depends on whether the integrals of the {following} form vanish:

    \begin{alignat}{2}
        I_{W', W''\to W}\propto \label{theI}
        \chi^{(2)}_{\alpha\beta\gamma} \int\limits_{V} \text{d} V \nonumber
        [\vec N_{\alpha}\cdot \vec W_{p_ip_rm n }(2\omega)]\\
        \times [\vec N_{\beta}\cdot \vec W_{p_i'p_r'm'n'}(\omega)]
        [\vec N_{\gamma}\cdot \vec W_{p_i''p_r''m'' n''}(\omega)]
    \end{alignat}

    Here $\vec N_{\alpha}$ are basis vectors of the Cartesian coordinate system which are proportional to electric vector harmonics with $n=1$,
    taken at zero frequency~\cite{bib:arxiv}, $\vec N_{x}\propto\vec W_{-1111}(0), \ \vec N_{y}\propto\vec W_{-1-111}(0), \ \vec N_{z}\propto\vec W_{-1101}(0)$.
    Note that summation is performed over repeated indices, and the integral~\eqref{theI} is the sum of the integrals of three scalar products of vector
    spherical harmonics over all nonzero $\hat\chi^{(2)}$-tensor components.

    %
    %

    First, we consider {the} integrals for {the} nonzero components of the $\hat\chi^{(2)}$-tensor. According to the selection rule theorem for matrix
    elements~\cite{bib:landau}, such {an} integral over the symmetric nanoparticle's volume will be
    nonzero only if the integrand contains an invariant with respect to all transformations of {the} particle{'s} symmetry.
    Since the integrand is a scalar value, it is convenient to expand it in a series in terms of scalar functions $\psi_{p_rmn}(\vec r, \omega)$:
    $\psi_{^{~1}_{-1}mn}(\vec r, \omega)=j_n(\sqrt{\varepsilon} \frac{\omega}c r){^{\cos m \phi }_{\sin m \phi }P_n^m (\cos \theta)}$.
    Their radial part is invariant with respect to any point transformations, and the angular part is transformed in a known way \cite{Zhang:08}.
    Examples of scalar functions that contain invariants with respect to transformations of different particl{e} symmetry groups are shown in Table 1 (first column).
    These functions contain invariants only if the particle is located in a certain way relative to the coordinate axes.
    For a particle rotated in an arbitrary fashion, they must be transformed with Wigner D-matrices.
    In order to understand whether any of the terms of the \blue{ integral \eqref{theI} contain} functions invariant with respect to particle symmetry
    transformations, we first decompose each of the scalar products in functions $\psi_{p_rmn}(\vec r, \omega)$ (with $|\vec r|$-dependent coefficients).
    Clebsch-Gordan expansions for scalar products of spherical vectors are known from the literature \cite{bib:VMH} (7.3.10).
    Based on them, we obtain the rules that determine {the} numbers $p_r$, $m$, $n$ \cite{bib:arxiv} (Appendix~B).
    These rules are based on the fact that the vector spherical harmonics with the numbers $m$, $n$ behave in the same way as scalar harmonics under rotations,
    {though,} {when reflected,} the behavior of the magnetic harmonics is opposite.
    It is also necessary to take into account the orthogonality properties of vector functions{,} such as: $[\vec N_{-10n'}\cdot \vec M_{10n}]=0$, $[\vec N_{p_rmn}\cdot \vec M_{-p_rmn}]=0$.
    After this procedure, only the sum of the products of three scalar functions remain {in} the integral $ I_{W', W''\to W}\propto \chi^{(2)}_{\alpha\beta\gamma} \int\limits_{V} \text{d} V
    [\sum_{ p_r, m, n}c(r)\psi_{p_rm n }]
    [\sum_{p_r', m', n'}c'(r)\psi_{p_r'm' n' }]\cdot
    [\sum_{p_r'', m'', n''}c''(r)\psi_{p_r''m'' n'' }]$,
    and they in turn can be decomposed again into scalar functions with known coefficients~\cite{bib:arxiv, bib:math}:
    $ I_{W', W''\to W}\propto \chi^{(2)}_{\alpha\beta\gamma} \int\limits_{V} \text{d} V
    [\sum_{ p_r, m, n}d(r)\psi_{p_rm n }]$.
    
      In the case of a spherical particle, the integral $ I_{W', W''\to W}$ {can} be nonzero only if the coefficient before the $\psi_{100}$ is nonzero.
    For particles of other symmetries, the presence of the spherically symmetric function $\psi_{100}$ is not necessary, the integral may be nonzero if the
    decomposition includes at least one of the functions $\psi_{p_rmn}$ which are invariant with respect to {the} symmetry transformations of the given particle.
    This changes the selection rules, for example, the "triangle rule" disappears for all particles except spherical {ones}.

\begin{figure}[h!]
    \begin{center}
        \includegraphics[width=0.4\linewidth]{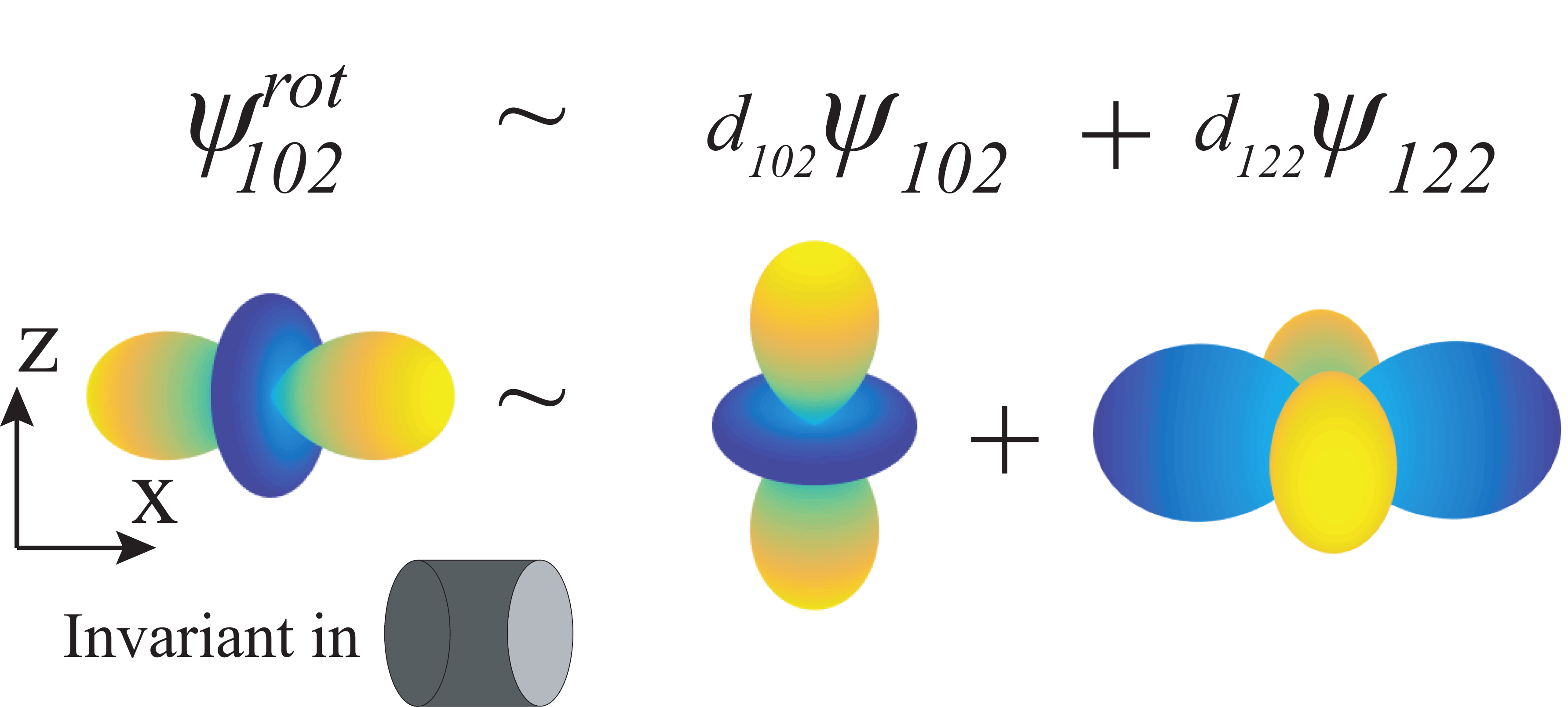}    \caption{Figure 1. Transformation of spherical function under rotation}
        \label{figrr}
        \end{center}
    \end{figure}

    Table~\ref{table1} provides simplified selection rules that do not take into account the possible orthogonality of vector functions.
    The product of the parities $p_i$ $p_r$ as well as the sum of the projections $m$ (the signs $\pm$ can be chosen arbitrarily) of all harmonics
    under the integral, should give the parities and projections of some functions which are invariant with respect to the particle's group transformations.
    Due to the form of the integral \eqref{theI}, the unit vectors of the Cartesian coordinate system have indices of vector spherical harmonics with $n=1$.
    The corresponding projections in the table are denoted as $m^{\alpha}$, $p_r^{\alpha}p_r^{\beta}p_r^{\gamma}= p_r^{\alpha\beta\gamma}$, $p_i^{\alpha}p_i^{\beta}p_i^{\gamma}= -1$.
    The table shows that the lower the symmetry of a particle, the weaker the selection rules for SHG. 
    However, it can be {seen} that for a cylinder lying on its side, the rules look "weaker" than for a cylinder oriented along the $ z$-axis. \blue{This is due to the fact that selection rules don't provide the exact form of the coefficients before each multipole.  For example, the function $\psi_{102}$, which is invariant for the $z$-oriented cylinder, is presented as a sum of the functions $\psi_{p_rm2}$ when rotated \cite{Zhang:08, Wunderlich2014}, so, for the arbitrarily oriented cylinder the specific combination of $\psi_{p_rm2}$ is invariant, while the selection rules can't specify it (see Fig.~\ref{figrr}). }
    {Because of this}, it is advisable to choose the most natural location of the coordinate system, which wil{l r}eveal a greater number of forbidden transitions. \blue{When we choose an arbitrary coordinate system, the multipoles with all possible $m$ and $p_r$ appear in a generated field, but they are still the multipoles with specific $m$ and $p_r$, but rotated with help of Wigner D-matrixes.  For a similar reason, selection rules at the bottom of table~\ref{table1} look different for different orientations of the prism or pyramid, but these are the same rules in different coordinate systems. }
    
    \blue{The SHG from structures with Td lattice ($\chi_{\alpha\beta\gamma}=\chi|\epsilon_{\alpha\beta\gamma}|, \ \epsilon - $ Levi-Chivita tensor), especially AlGaAs or GaAs nanostructures, is extensively studied \cite{KrukSS, Gili:16, CMalga}, and interesting properties of the SH signal revealed \cite{ Ghirardini:17, 7911300}, so, on the right side of table~\ref{table1} we illustrate the selection rules with the example of such structures. We depict the radiation patterns of allowed multipoles generated by the $x$-polarized plane wave, whose $\vec E$-field is even under $y=0$-plane reflection ($p_r'= p_r''=1$), and $m'^{\omega}=m''^{\omega} =1$ \cite{bib:Boren}. Crystalline axes $[001], [010], [100]$ are oriented along the coordinate system axes, so in such material $p_r^{\alpha\beta\gamma}=-1$, $\pm m^{\alpha}\pm m^{\beta}\pm m^{\gamma}=0, \pm 2$. Applying the rules for a $z$-oriented cylinder, we should take $p_r=-1$ and $m^{2\omega}=0,2,4$ to get the invariant functions $\psi_{10(2z)}$ under the integral \eqref{theI}. Radiation patterns of some multipoles with these numbers are provided in the table. When we reduce the symmetry of a structure, additional multipoles become allowed, examples of their appearance are shown in comparison with a higher-symmetry cylindrical structure. The plane wave has low symmetry itself, so the differences between different structures are not very strong, sometimes appearing only for high multipolar orders. However, the contributions of the multipoles are defined by the resonances of the structure, and sometimes it's reasonable to consider only major contributions to the fundamental wave, which makes the rules more specific. We consider an example of a single-mode excitation in the next section.
    Note that if we change the orientation of the crystalline lattice or excitation wave, the corresponding indexes will change and it is necessary to apply the rules again.  }

    Although the rule{s g}iven in the table allow a large number of forbidden transitions to be detected, sometimes it is necessary to decompose the integrand over scalar
    functions for each of the $\hat \chi$-component, as described in \cite {bib:arxiv}, to obtain more strict selection rules.
    This still does not exclude the existence of specific prohibitions, for example, due to integration over the radius.
    However, if we pay attention to the properties of the tensor $\hat\chi^{(2)}$ during the crystalline lattice's group transformations, we can extract additional
    information about the selection rules in the system. So far, the obtained rules applied to each of the addends in \eqref{theI} separately, however, after summing
    up over the repeated indices $\alpha, \beta, \gamma$, additional restrictions may appear due to the fact that the addends mutually destroy each other because of lattice symmetry.
    The method of detecting these prohibitions is based on the fact that under symmetry transformations, which are common elements of particle's and lattice's groups,
    (the mutual orientation of the lattice and the particle is important) the {intergral} \eqref{theI} must be invariant as a whole.  Indeed, under transformations from the crystalline lattice's group, the tensor $\hat \chi^{(2)}$ is not transformed, and only three vector harmonics are transformed, which means that for the
    integral over the particle volume to be nonzero, the product $\vec W_{p_i''p_r''m'' n''}(\omega) \otimes \vec W_{p_i'p_r'm' n'}(\omega) \otimes \vec W_{p_ip_rmn}(2\omega)$
    should contain an invariant for all particle transformations that do not change the tensor $\hat \chi^{(2)}$.
    Thus, it is necessary to consider the behavior of vector harmonics with respect to the intersection of the particle's and lattice's groups, and to determine whether their product contains an invariant representation.
    Symmetry classification and irreducible representations of vector spherical harmonics for finite groups are given, for example, in \cite{bib:arxiv, bib:arxiv2}.
\section{\blue{Single-mode excitation}}
        \begin{figure}
        \begin{center}
            \includegraphics[width=0.9\linewidth]{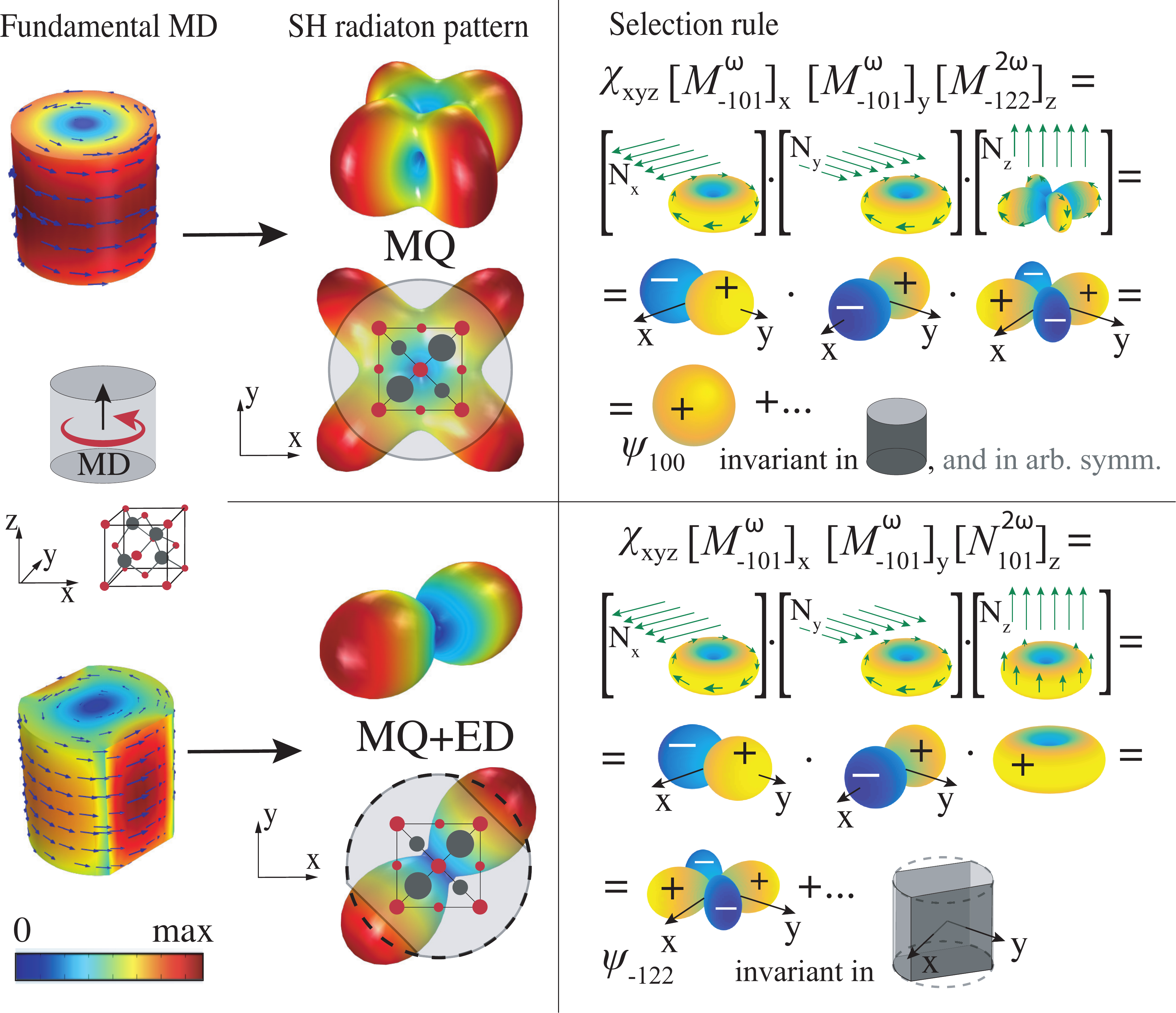}
            \caption{Figure 2. Second-harmonic radiation patterns generated by a magnetic dipole (MD) at the fundamental frequency for the GaAs cylinder \blue{(top left)},
                and the GaAs cylinder truncated laterally along the plane $x~=~y$ \blue{(bottom left)}. For the cylinder, the main generated mode is the magnetic quadrupole (MQ).
                In the truncated cylinder the electric dipole (ED) is also generated, which creates a radiation pattern directed along the $x = y$ axis when
                interfering with the magnetic quadrupole. \blue{The relative position of the $T_\text{d}$-lattice axes is shown}.
                The cylinder's radius is 140~nm, height is 280~nm, the wavelength is 1480~nm (for $\omega $) and 740~nm (for $ 2 \omega $), $ \varepsilon = 12.96 $.   {The width} of the truncated cylinder is 230~nm. \blue{On the right side the illustration of derivation of the selection rules given by the integral \eqref{theI} is shown. Cartesian projections of vector spherical harmonics $[\vec N_{\alpha}\cdot \vec W ]$ are the scalar functions with particular symmetry which give invariant functions after multiplication.}}
               \label{fig1}
        \end{center}
    \end{figure}
    
  According to \cite{bib:MD, bib:MD2}, selective excitation of individual multipoles at the fundamental frequency is possible.
    Therefore, for simplicity, we consider SHG by {a} magnetic dipole $\vec M_{-101}$ (parallel to the $z$-axis) in {a} monocrystalline nanocylinder (Fig.~\ref{fig1}) from {gallium arsenide}
 ($T_\text{d}$ lattice, $\varepsilon = 12.96$, {with} the crystal axe{s d}irected along the coordinate axes).
    {A s}imilar geometry was {studied} experimentally in \cite{bib:cylexp}. To study the effect of symmetry breaking, a cylinder truncated along the $x = y$ plane was also considered.
    The wavelength corresponding to the fundamental frequency $\omega$ is 1480~nm, the cylinder{'s} radius is 140~nm, and {it's} height is 280~nm.
    The smallest width of the truncated cylinder is 230~nm. We computed the SH radiation patterns with COMSOL Multiphysics for both nanoparticles{. The calculations show} that for the whole cylinder the main share of the radiation is accounted for the magnetic quadrupole ($\vec M_{-122}$), and negligibly small for higher-order multipoles, since the frequency is far from their resonances.
    In the case of symmetry breaking ({truncated cylinder}), the generation of the electric dipole along the $z$-axis ($\vec N_{101}$) {is} allowed, and the main part of the
    radiation energy falls on this mode, which interferes with the magnetic quadrupole, which is reflected in the radiation pattern. 

    Applying the selection rules to SHG from a magnetic dipole in the cylinder with $T_\text{d}$ lattice we obtain that the only non-zero integrals
    (here we recall the orthogonality properties of vector functions) are
    $ \chi^{(2)}_{zxy} \int\limits_{V} \text{d} V
    [ W_{-1-12n}(2\omega)]_z
    [M_{-101}(\omega)]_x
    [M_{-101}(\omega)]_y\blue{\propto \chi^{(2)}_{zxy} \int\limits_{V} \text{d} V
    [ \psi_{-12(2z)}]
    [\psi_{-111}]
    [\psi_{111}] }$, and for large wavelengths, the main contribution is from $\vec W_{-1-122}=\vec M_{-122}$ \blue{($[\vec M_{-122}]_z=\psi_{-122}$). This integral is illustrated on the right side of fig.~\ref{fig1}.}
    In order to allow generation of the electric dipole along $ z $, the integral
    $ \int\limits_{V} \text{d} V
    [ N_{101}(2\omega)]_z
    [M_{-101}(\omega)]_x
    [M_{-101}(\omega)]_y\propto \int\limits_{V} \text{d} V( c(r) \psi_{-122}+c'(r) \psi_{-124})$\blue{, which is also illustrated on the right side of fig.~\ref{fig1},}  should contain {an} invariant, to achieve this, we break the
    symmetry of the particle by making small cuts along the x = y direction.
    Since the effect of the appearance of an electric dipole is in violation of symmetry, and it is not associated with resonances, it manifests itself in a range of wavelengths.
    Moreover, by changing th{e w}idth of the truncated cylinder, it is possible to control the value of the contribution of the electric dipole.
    By breaking symmetry along other directions, it is also possible to allow the generation of other dipole (and higher) modes.
    Note that such cuts also affect the fundamental mode by adding {other} multipoles to it, but this effect is
    insignificant and does not affect the selection rules due to the fact that the admixed multipoles have a certain symmetry.

    %
    %

    {\bf Conclusion}.
    The paper explains the effect of the nanoparticle shape on the second-harmonic signal, and presents a method for finding the selection rules for particles of
    arbitrary shape{s} from any non-centrosymmetric material. A table of selection rules for particles of the most common forms is given. The obtained
    selection rules {are} also be suitable for other nonlinear processes, such as spontaneous parametric scattering, generation of higher harmonics, sum-frequency, etc.,
    with minor changes. Due to the fact that even {a} slightly imperfect particle shape can {significantly} affect the generated fields,
    the results of {this} work can {be used to} explain the possible discrepancies between theoretical predictions and experimental observations.

    %
    %
\section*{Acknowledgments}
    The authors thank Petrov M.I., Poddubny A.N., Toftul I.D., Koshelev K.L., Nikolaeva A.A. for the fruitful discussions. The work is supported by the Russian Science Foundation (grant 18-72-10140).



\bibliography{liter_2}






\end{document}